\documentclass[12pt,notitlepage,onecolumn]{article}

\usepackage{amsmath}
\usepackage{amssymb}
\usepackage{bm}
\usepackage{graphicx}
\usepackage{amsbsy}
\usepackage{color}

\begin{document}

%\begin{document}
%\vspace*{0.2in}

%\begin{flushleft}
 \begin{center}
{\Large
\textbf\newline{SME investment best strategies  \\  Outliers for assessing how to optimize performance.    
% \\ An answer from extreme value statistics.
} % Please use "title case" (capitalize all terms in the title except conjunctions, prepositions, and articles).
}
\newline
\\
  Marcel Ausloos \textsuperscript{1,2,*} ,  \\  Roy Cerqueti \textsuperscript{3},
  Francesca Bartolacci$^{4}$,  Nicola G. Castellano$^{5}$ 
 \\
\bigskip
 \textbf{$^1$} School of Business, University of Leicester, University Road. \\Leicester, LE1 7RH, United Kingdom. \\Email: ma683@le.ac.uk\\
 \textbf{$^2$}  GRAPES -- Group of Researchers for Applications of Physics in Economy and Sociology. Rue de la Belle Jardini\`ere 483, B-4031, Angleur, Belgium. \\Email: marcel.ausloos@ulg.ac.be
 \\
 %* Corresponding author \\ 
 \textbf{$^3$} Department of Economics and Law, University of Macerata, \\Via Crescimbeni 20, I-62100, Macerata, Italy.
%Tel.: +39 0733 258 3246; Fax: +39 0733 258 3205.\\
 \\Email: roy.cerqueti@unimc.it.
 \\
 \textbf{$^4$} Department of Economics and Law, University of Macerata, \\Via Crescimbeni 20, I-62100, Macerata, Italy.
%Tel.: +39 0733 258 3246; Fax: +39 0733 258 3205.\\
 \\Email:  bartolacci@unimc.it
 \\
  \textbf{$^5$} Department of Economics and Law, University of Macerata, \\Via Crescimbeni 20, I-62100, Macerata, Italy.
%Tel.: +39 0733 258 3246; Fax: +39 0733 258 3205.\\
 \\Email: ncaste@unimc.it
\bigskip
%$\natural$ These authors contributed equally to this work.

 %$\natural$April 28, 2017

% Use the asterisk to denote corresponding authorship and provide email address in note below.

%JEL Classification numbers: C3, C12,  C14,  C55,  D31,  E01,  H24, J19
 \end{center}
%\end{flushleft}

% PACS: ???

%\clearpage
\begin{abstract}
Any research on strategies for reaching business excellence aims at revealing the appropriate course of actions 
any executive should consider. Thus, discussions take place on how effective a performance measurement system can be 
estimated, or/and validated. Can one find an adequate measure (i) on the performance result 
due to whatever level of investment, and (ii) on the timing of such investments? 
We argue that extreme value statistics provide the answer. We demonstrate that the 
level and timing of investments allow to be forecasting small and medium size enterprises (SME) performance,
 - at financial crisis times. The "investment level" is taken as the yearly total tangible asset (TTA). 
 The financial/economic performance indicators defining "growth” are the sales or total assets variations; 
  "profitability" is defined from returns on investments or returns on sales. 
  Companies on the Italian Stock Exchange STAR Market serve as example.  
  It is found from the distributions extreme values that outlier companies (with positive performance) 
  are those with the lowest but growing TTA. In contrast, the SME with low TTA, 
  but which did not increase its TTA, before the crisis, became a "negative outlier". 
   The outcome of these statistical findings should suggest strategies to SME board members.
\end{abstract}
%keywords : distribution outliers; business performance; small and medium size enterprises; total tangible assets;
%Word count :  $\sim  8000$
% \pacs{89.75.Fb, 89.75.Da, 89.75.Kd,05.65.+b, 02.50.-r}
  % \maketitle
%\section{Foreword, - to be removed}\label{forw}
%This short note can be considered as a comment, nevertheless with some rather original data and findings, on  recent considerations, about  an analytical relationship between the kurtosis and the skewness of a data distribution, which "culminate" in a recent paper by Cristelli et al. Indeed, some time ago Cristelli et al. presented an intriguing relationship  as being a  "universal relation(s) between skewness S and kurtosis  K in complex dynamics": $K\simeq S^{\nu}$, with $\nu= 4/3$.  \cite{CristelliPRE,cristelli2014universal}
%K  N1/3S4/3

%   \textcolor{red}{ strength} \textcolor{green}{ length}
\section{Introduction}\label{Introduction}
The statistics literature is filled with hundreds of papers on how to apply methods in order to measure, assess, discuss the reliability of  financial data, (Amendola et al., 2006; Amendola et al., 2008; Boente et al., 2010) and develop  strategies or  suggest hints toward "better performance".
Indeed, any research on strategies for reaching business excellence under an economic crisis aims at revealing the appropriate course of actions which any executive should consider, under the limitations and particular conditions that arise in an economic crisis environment (Afthonidis, and Tsiotras, 2014). The first reaction is often to implement cost saving policies, to interrupt investment plans and proceed to business restructuring with cost cutting in mind. This may have some direct results, yet will not secure the future of the enterprise (Koksal and Ozgul, 2007). Others consider that the first step of the management should be to secure adequate resources, especially liquidity, - but again, history has shown that, very rarely during a recession, has this defensive strategy brought satisfactory results in the long run (Reeves and Deimler,  2009).

Nevertheless, justifying an investment can be frustratingly difficult to suggest when the payback is measured by loosely convincing projected long-term increases in sales, assets, and other profitability performance measures. Thus, discussions take place on how effective a performance measurement system can be estimated, validated, or credible (Vitale and Mavrinac, 1995), - a huge statistical set of criteria! 

Thus, despite the episodic pervasiveness of recessions and their destructive impact on firms, and such mentioned considerations, a void exists in the management literature examining the intersection between recessions, strategy, and performance (Behrens et al., 2004; Latham and  Braun, 2011). In fact, it can be asked whether there is an "initial condition" which in this non-linear set of plans and activities determines, and the more so allow to forecast, the future performance; in other words (Bourne  and Neely,   2001) one has to wonder why measurement initiatives succeed and fail.  The more so, one should avoid  extreme losses (Vaz de Melo Mendes,  2006) and  rather aim at huge gains based on some {\it ad hoc} strategy (Roberts,  2003; Reeves and  Deimler,  2009). 

For statistical purposes and reasoning, can one find an adequate measure (i) on the performance result due to whatever level of investment, and (ii) on the timing of such investments? Is it simply obvious that a control of the investment strategy will lead to an improved (hopefully optimal) performance? Some answers to these questions are the aims of this report.

The information and feedback from the measures should be used to challenge the assumptions and test the validity of whatever strategy (Eccles and Pyburn, 1992; Kaplan and Norton, 1996; Feurer and Chaharbaghi, 1995). In fact, authors have argued that they should be used for both purposes (Feurer and Chaharbaghi, 1995; Kaplan and Norton, 1996). Therefore, assessing the implementation of strategy and challenging the strategic assumptions are the two main subdivisions of the use of the performance measures.  On performance measure suggesting investment policies, at a time of crisis, let us not go back too far  recalling history; select a recent one "THE crisis", for   immediately connecting thoughts to the above questions.  Let us  rather consider the practical side of the investigation: the use of  statistics in performance measures, allowing for survival (Datta et al., 2016).

Focusing on SME, an increase performance may depend on very appropriate investment strategies, more than for multinational firms. Innovation input and development are surely traditional set-ups, but others can arise from "more internally based" pertinent strategies imagined by the executive board. %They can be sometimes audacious (Kailasam et al.,  2016).

%Neely  and  Szwejczewski (2001) have studied the performance of SMEs. 
Neely has much elaborated  on measuring operations performance of SMEs, - alone or with coworkers (Nellly, 1997; Nelly and Austin, 2002; Kennerley  and  Neely,  2002; Neely and  Al Najjar,  2006), distinguishing various points of view and evaluating relevant filters for analysis, even employee and/or customer satisfactions. For  completeness, and in view of the specificity of the report, using Italy stock market as the case, let us point also (because of the specificity of the study,  the Veneto Region in Italy and the East of England, UK) to Neely et al. (2001) on the impact of innovation on business performance.

Thus,  recalling such works incites to consider "extreme profitability" as another measure beside "huge growth" among the indicators of interest. Whence, we should  obtain some statistical inference using extreme order statistics (Caroni and Karioti, 2004; Unnikrishnan, 2010; Gumedze and Chatora, 2014).
%Pickands,  1975).
 This is developed in Sections \ref{basicRQ}-\ref{Resanalysis}.
In Section \ref{conclusion},  with some summary, we offer some conclusive remarks and provide also suggestions for future research directions.
%Appendix B exhibits a condensed matter argument, in order to provide some physical intuition for  the rank-size analysis parameters, whence of the $K-S$ relationship. 
%

\section{ Thus, the most basic  economic strategy questions  tied to statistics are}\label{basicRQ}

"Despite the episodic pervasiveness of recessions and their destructive impact on firms, a void exists in the management literature examining the intersection between recessions, strategy, and performance" wrote Latham  and Braun (2011).  Our paper seeks to address and bridge this research gap, finding a statistical relationship  between marketing strategies and performance $ca.$ an economic crisis.  Such a focussed  aim is also raised by Koksal and Ozgul  (2007).    

 It seems that our common agreement on the most basic or pertinent questions is about the levels of investments that can be used, - and in what timing order. Obviously, one needs to rely on a preliminary acceptable measure of the so called "investment efficiency." 
Thereafter, the questions appear to be two fold, - with respect to the quantitative aspects: (1) Should one (later) measure the efficiency in terms of the lowest investment, or (2){\it a contrario}, is a high amount of investment necessarily for   a better performance? In fact, it can be hard to decide  what  consecutive investments (one "low" followed by one "high", or the other way around, or even with some longer cycling) is responsible for a subsequent efficiency.

The main point (H1) to be clarified pertains of course to the  (statistics) definition of the "business performance efficiency" measure.  It is here considered that only a few aspects seem relevant. In view of the  pertinent literature, we select four variables, or financial/economic indicators, for representing business performance: two of them for "growth", which can be expressed through (i) sales variations (DS) and (ii) total assets variations (DA), and also  two for "profitability",  through (iii) returns on investments (ROI) and (iv) returns on sales (ROS).

Next, (H2) one can admit that a certain time span has to be used for obtaining a reliable measure.  These indicators will be measured  here from publicly available results (in  2008, 2009, and  2010), AFTER the crisis, averaged over such a 3 year time interval: the notation will be for example $<DS >_3$ for the sales  ($S$) variations, averaged over 3 years: [2008-2010]. The variable of interest measuring some level of investments is taken to be the firm total tangible assets (TTA). The data of interest BEFORE the ("unknown" or incoming) crisis is chosen to cover 2006 and 2007. It will be noted as TTA06 or TTA07; moreover, its average is noted $< TTA >_2$.

\subsection{Statistical analysis methodology}\label{statanal}

After having performed the 3 year averaging for (i)-(iv), the methodology goes as follows: each (i)-(iv) average values are used as the numerator of the "performance efficiency" ratio in which the denominator is either the lowest TTA (TTAm) or the highest TTA (TTAM), value in either 2006 or 2007. Thereafter, the (i)-(iv) averages of the firms are also compared with respect to the TTA average trough their ratio for which the denominator is $< TTA >_2$, identical to (1/2)(TTA06 + TTA07), of course. This leads to 12 indicators. The (62, at that time) SME on the STAR Market Segment of the Italian Stock Exchange are considered to span various types of SME and a convenient sample for examining statistical characteristics leading to conclusions on performance efficiency. N.B. The STAR (Segment for High Requirement Shares) market

\noindent 
 $http://www:borsaitaliana:it/azioni/mercati/star/home-star/segmento-star:en.htm$ 
\noindent 
 includes companies capitalized from 40 million to 100 million Euros; see: 
\noindent 
$http://www:borsaitaliana.it/homepage/homepage.htm$
\noindent 
within the Milano electronic share market (Mercato Telematico Azionario: MTA):
\noindent 
$http://www:borsaitaliana:it/azioni/mercati/mta/home/mta-mercato-telematico-azionario:en.htm$

\subsection{Stressing the usefulness of extreme values}\label{extremevalues}
A very fundamental point is next emphasized: it should be easily understood and accepted that the statistical outliers are the companies giving a better view of the success or failure of their previous investment strategy. The outliers overperform or underperform. That is what is usually to be avoided or searched for, whence to be attracting the discussion.: means are often considered. However such values, whence firms, for which the final outcome occurs "near the average" are in fact "strategically uninteresting", - because merely falling within statistical error bars; thus, they should not be considered to be relevant for our purposes. Therefore, the outliers are next extracted, shone upon, and discussed for emphasizing the interesting features allowing recommendations.

This reasoning is in line with the statistical literature which includes work on exploring possible trends in damages resulting from extreme events, like %landfalling, due to hurricanes (Xiao et al., 2015) % or tsunamis, due to evolving climate conditions (Sharma et al., 2012)
 earthquakes (Pisarenko and Sornette, 2003; Sornette and Werner, 2011) or floods (Akinsete et al., 2008), and survival analysis (Datta et al.,  2016).

   \begin{table} \begin{center}
\begin{tabular}[t]{|c||c|c|c|c|c|c|c|}
  \hline
%& \multicolumn{2}{|c|}{ $ATI_{c,p}$}&  \multicolumn{2}{|c|}{ $N_{inhab,c,p}$}  \\ \hline
 
Variable	&	Min.	&	Max.	&	Sum	&	Mean 	&	StDev 	&	Skewness	&	Kurtosis	\\	
 	&		&	 	&		&	($\mu$)	&	($\sigma$)	&		&		 	\\\hline	
$ TTAm$	&	42.000&	4.829 10$^5$&	2.600 10$^6$ &	41931	&	89262	&	3.4022&	11.948	\\
$TTAM$	&	131.00&	5.321 10$^5$ &2.893 10$^6$ &	46662	&	96049	&	3.3905&	12.156	\\
$<TTA>_2$	&	86.5	&	5.075 10$^5$	&	2.746 10$^6$	&	44297	&	92600 	&	3.3967	&	12.062	\\\hline		
$<DS>_3$	&	-0.1924	&	1.1767	&	4.9303	&	0.0795	&	0.198	&	3.1414	&	14.013	\\	
$<DA>_3$&	-0.1436	&	1.9818	&	7.8786	&	0.1271	&	0.330	&	3.8060	&	16.885	\\	
$<ROI>_3$&	-0.0768	&	0.3457	&	3.0115	&	0.0486	&	0.067	&	1.5342	&	5.1206	\\	
$<ROS>_3$&	-0.6609	&	0.2445	&	2.5316	&	0.0408	&	0.116	&	-3.505	&	20.046	\\\hline	
 \end{tabular}
   \caption{     Summary of  (rounded) statistical characteristics  for  the time average distributions of the growth and profitability indicators for the 62 STAR companies, and of their $< TTA >_2$ , in the center of the table, in per cents and in $10^6$ Euros, respectively; the skewness and kurtosis are dimensionless scalars.}\label{TablestatSKSK}
\end{center} \end{table}

   \begin{table} \begin{center}
\begin{tabular}[t]{|c||c|c|c|c|c|c|c|}
  \hline
%& \multicolumn{2}{|c|}{ $ATI_{c,p}$}&  \multicolumn{2}{|c|}{ $N_{inhab,c,p}$}  \\ \hline
 
Variable	&	Min.	&	Max.	&	Sum	&	Mean 	&	StDev 	&	Skewness	&	Kurtosis	\\	
 	&		&	 	&		&	($\mu$)	&	($\sigma$)	&		&		 	\\\hline	
$< DS >_3$ / $TTAm$&  	-0.01482	&	0.4795	&	1.1602	&	0.018.71 	&	0.08278	&	5.1151	&	24.804	\\
$<DA >_3$/$TTAm$	& -0.11547	&	0.5089 	&	0.4468	&	7.20  10$^{-3}$	&	0.06747 	&	6.6860	&	48.151	\\
$<ROI >_3$ /$TTAm$& -0.01313	&	0.1573	&	0.4007 	&	6.46  10$^{-3}$	&	0.02612	&	4.8640	&	23.330	\\
$< ROS >_3$ /$TTAm$& -0.24661	&	0.1623 	&	0.1533	&	2.47  10$^{-3}$	&	0.04138 	&	-2.2285	&	23.790	\\ \hline
$< DS >_3$ /$TTAM$& -8.91  10$^{-3}$	&	0.3962  	&	0.6726 	&	0.01085	&	0.05379 	&	6.3815	&	41.721	\\
$< DA >_3$ /$TTAM$& -0.03702 	&	0.4524 	&	0.4827 	&	7.79 10$^{-3}$	&	0.05810	&	7.3938	&	54.155	\\
$<ROI >_3$ /$TTAM$& -6.53 10$^{-3}$	&	0.0733  	&	0.1894	&	3.06 10$^{-3}$	&	0.01127 	&	4.8538	&	24.937	\\
$<ROS >_3$ /$TTAM$& -0.1226 	&	0.0573 	&	0.0748 	&	1.21 10$^{-3}$	&	0.19334	&	-3.5113	&	27.249	\\ \hline
$<DS >_3$/$< TTA>_2$	& -9.70   10$^{-3}$	&	0.4195 	&	0.8094	&	0.01306	&	0.0607 	&	5.7396	&	33.346	\\
$<DA >_3$/$<TTA>_2$	& -0.05607	&	0.4790 	&	0.4882 	&	7.87  10$^{-3}$	&	0.0619 &	7.2609	&	52.930	\\
$<ROI>_3$ /$<TTA>_2$	& -8.72  10$^{-3}$	&	0.1000 	&	0.2479 	&	4.00 10$^{-3}$	&	0.0154 	&	4.9740	&	25.738	\\
$<ROS>_3$/$<TTA>_2$	& -0.1638 	&	0.0788 	&	0.0900 	&	1.45  10$^{-3}$	&	0.0260	&	-3.3352	&	26.988	\\ \hline
	
 \end{tabular}
   \caption{ Summary of  (rounded) statistical characteristics  for  the  12 statistical indicators distributions of the growth and profitability measures for the 62 STAR companies;   in the center of the table, data is given in per cents;  the skewness and kurtosis are dimensionless scalars.
   }\label{NewTablestatSKSK}
\end{center} \end{table}

   \begin{table} \begin{center}
\begin{tabular}[t]{|c||c|c|c|c|c|c|c|}
  \hline
%& \multicolumn{2}{|c|}{ $ATI_{c,p}$}&  \multicolumn{2}{|c|}{ $N_{inhab,c,p}$}  \\ \hline
 
Indicator	& Mean 	&	StDev 	&$\mu-2\sigma$	&$\mu+2\sigma$\\	
 	&	($\mu$)	&	($\sigma$)	&		&		 	\\\hline	
$< DS >_3$ / $TTAm$&  1.8713	10$^{-2}$&	0.082777	&	-0.14684	&	0.18427	\\					
$<DA >_3$/$TTAm$	& 7.2064  10$^{-3}$&	0.067471	&	-0.12774	&	0.14215	\\					
$<ROI >_3$ /$TTAm$& 6.4631  10$^{-3}$	&	0.026115	&	-0.045767	&	0.058693	\\					
$< ROS >_3$ /$TTAm$& 2.4721  10$^{-3}$	&	0.041382	&	-0.080291	&	0.085235	\\ \hline					
$< DS >_3$ /$TTAM$& 1.0849	10$^{-2}$&	0.053792	&	-0.096734	&	0.11843	\\					
$< DA >_3$ /$TTAM$& 7.7854  10$^{-3}$	&	0.058099	&	-0.10841	&	0.12398	\\					
$<ROI >_3$ /$TTAM$& 3.0546  10$^{-3}$	&	0.011271	&	-0.019488	&	0.025597	\\					
$<ROS >_3$ /$TTAM$& 1.2058  10$^{-3}$	&	0.019334	&	-0.037463	&	0.039874	\\ \hline					
$<DS >_3$/$< TTA>_2$	& 1.3055 10$^{-2}$	&	0.060710	&	-0.10836	&	0.13447	\\					
$<DA >_3$/$<TTA>_2$	& 7.8741  10$^{-3}$	&	0.061904	&	-0.11593	&	0.13168	\\					
$<ROI>_3$ /$<TTA>_2$	& 3.9985  10$^{-3}$	&	0.015403	&	-0.026808	&	0.034805	\\					
$<ROS>_3$/$<TTA>_2$	& 1.4520  10$^{-3}$	&	0.025969	&	-0.050486	&	0.053390	\\ \hline					
 \end{tabular}
 \caption{ Indicators confidence interval limits (in per cents).    }\label{TableCI}
\end{center} \end{table}

  \begin{table} \begin{center}
\begin{tabular}[t]{|c||c|c|c||c|}
  \hline
 &  \multicolumn{4}{|c|}{Company Name}   \\ \hline
  efficiency&      (11) &  (13) Cairo&  (58) & (45) \\
  Indicator ratio&Buongiorno & Communication& Ternienergia& Mondo TV\\\hline
$< DS >_3$ / $TTAm$	&	0.4795	&	(0.0186)	&	0.4457	&	(0.0769)	\\
$<DA >_3$/$TTAm$	&	(-0.1155)	&	(-0.0217)	&	0.5089	&	(-0.0536)	\\
$<ROI >_3$ /$TTAm$	&	0.1277	&	0.1573	&	(0.0345)	&	(-0.0130)	\\
$< ROS >_3$ /$TTAm$	&	0.1623	&	0.1228	&	(0.0436)	&	-0.2466	\\\hline
$< DS >_3$ /$TTAM$&	0.1537	&	(0.0087)	&	0.3962	&	(0.0382)	\\
$< DA >_3$ /$TTAM$	&	(-0.0370)	&	(-0.0101)	&	0.4524	&	(-0.0266)	\\
$<ROI >_3$ /$TTAM$&	0.0409	&	0.0733	&	0.0306	&	(-0.0065)	\\
$<ROS >_3$ /$TTAM$	&	0.0520	&	0.0573	&	(0.0388)	&	(-0.1226 )	\\\hline
$<DS >_3$/$< TTA>_2$	&	0.2328	&	(0.0118)	&	0.4195	&	(0.0511)	\\
$<DA >_3$/$<TTA>_2$	&	(-0.0561)	&	(-0.0138)	&	0.4790	&	(-0.0356)	\\
$<ROI>_3$ /$<TTA>_2$	&	0.0620	&	0.1000	&	(0.0324)	&	(-0.0872)	\\
$<ROS>_3$/$<TTA>_2$	&	0.0788	&	0.0781	&	(0.0410)	&	(-0.1638)	\\\hline
\end{tabular}
   \caption{ Main positive and negative outliers of the growth variations and profitability efficiency indicators for the 62 STAR companies in per cent, i.e. those falling outside the interval $]$$\mu-2\sigma$, $\mu+2\sigma$$[$  corresponding to each ratio distribution. The data in parentheses correspond to those companies which are not truly outliers in a statistical sense for the index of interest, - “but almost”, like the inefficient Mondo TV.}\label{Table4outliers}
\end{center} \end{table}

   \begin{figure}   %1
    \begin{center}%centering
\includegraphics[scale=.65] {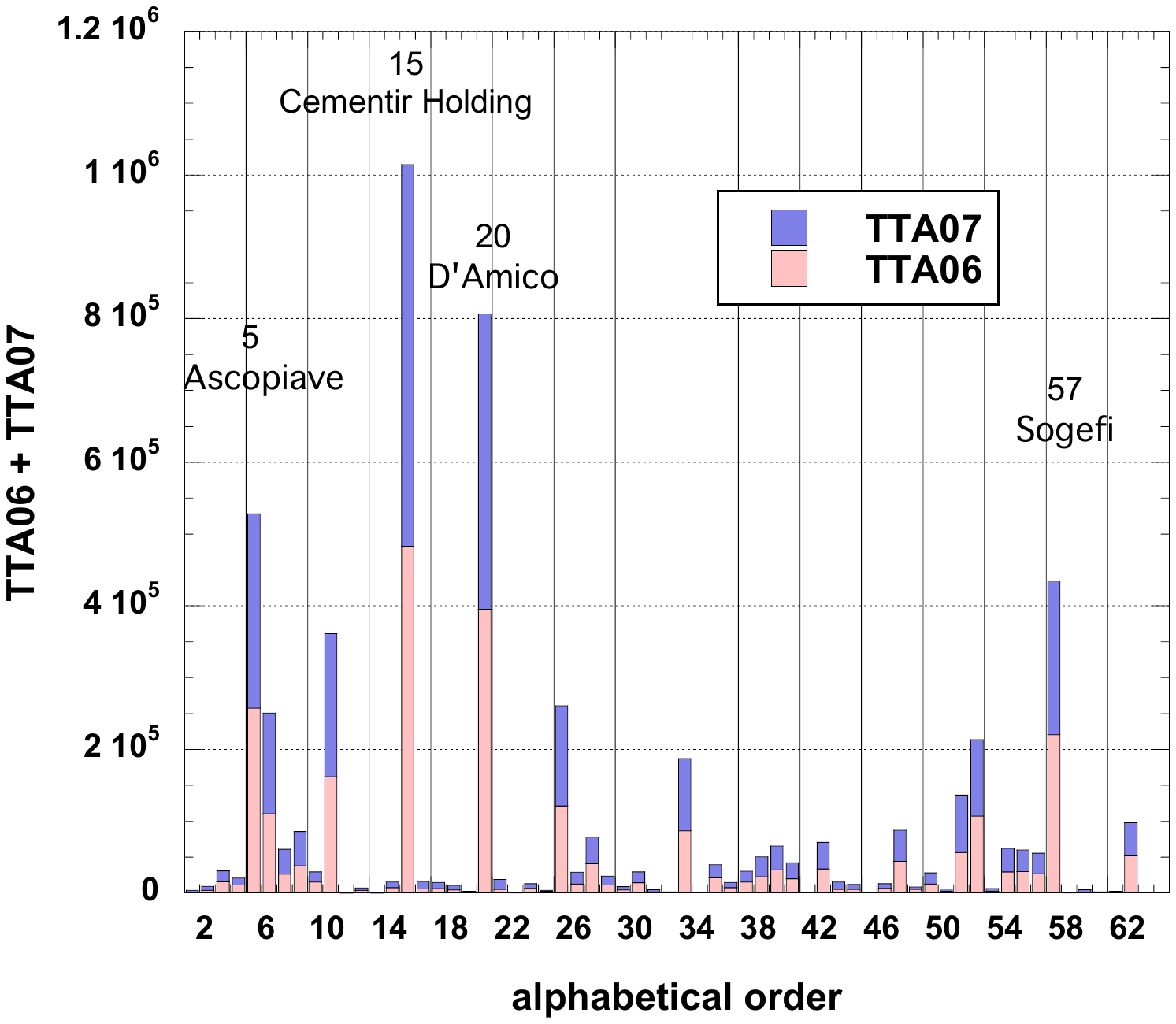}  
\caption{  Stacked $ TTA06$ and $ TTA07$  of  the 62 companies listed on the STAR market.  
 $x$-axis is the alphabetical index for the 62 SMEs.$
`$-. The four largest TTA firms are indicated. } 
\label{Plot30TTA06TTA07best.pdf}
\end{center}
\end{figure}

      \begin{figure} %2
  \begin{center}%centering
\includegraphics[scale=.65] {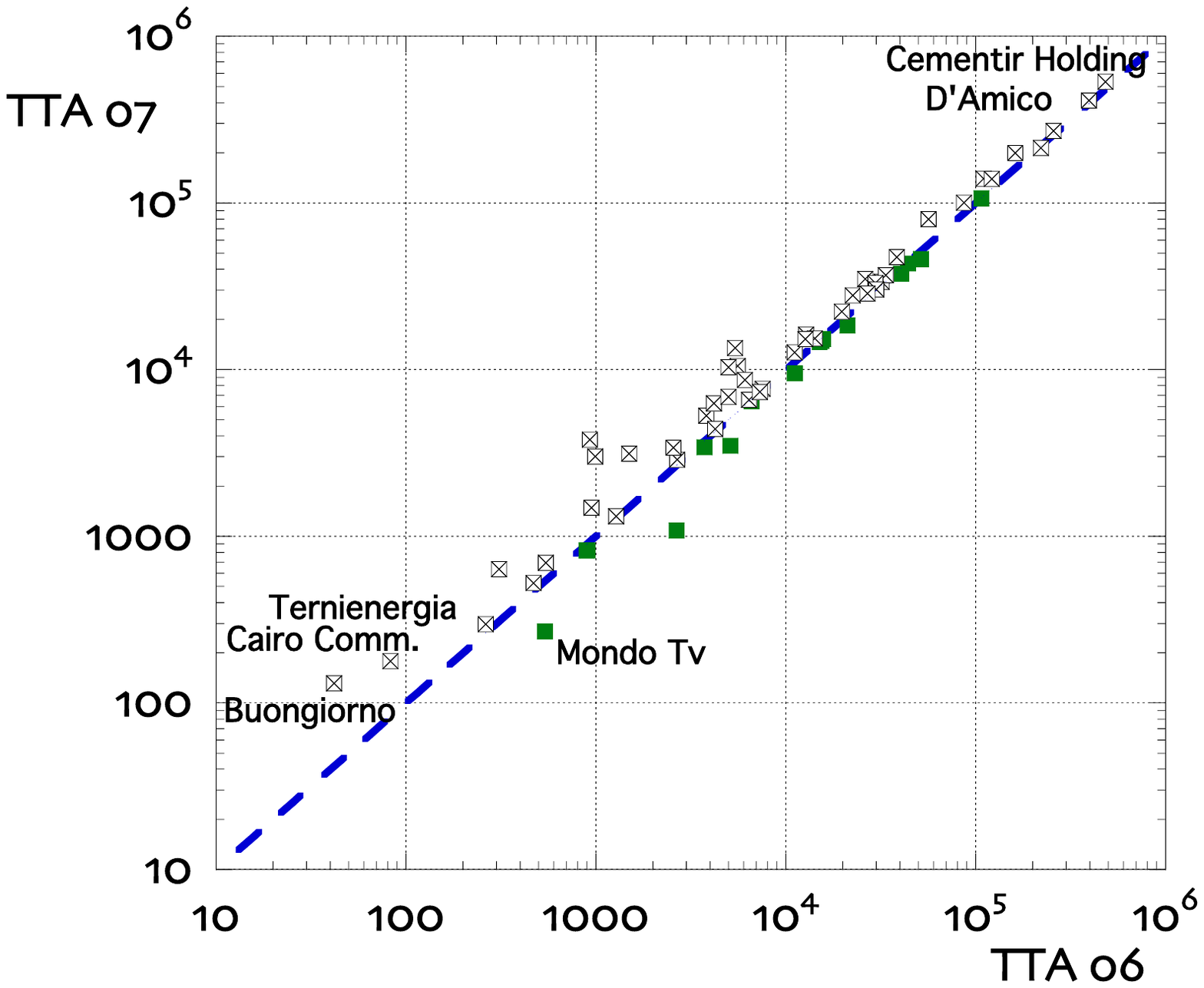}
\caption{  Relation   TTA07   $vs.$   TTA06   (in EUR)   for the 62 companies listed on the STAR market,  
distinguishing between  those with increased (open square with cross)  or decreased (full square) TTA. 
The name  of  a few "interesting", thus "extreme", companies is  pointed out.} 
\label{Plot1TTA06TTA07}
\end{center}
\end{figure}

  \section{Results  analysis}\label{Resanalysis}
 The raw data main statistical characteristics are given in Table \ref{TablestatSKSK}. Observe that since there is a negative minimum for each (i)-(iv) measure, some board  strategies were rather failures. Nevertheless, the mean is always positive. The distributions are quite extended, as indicated by the (easily estimated from the data in the table) so called coefficient of variation $\sigma/\mu$ values. The kurtosis is always positive and large, indicating lesser chances of extreme negative outcomes; the skewness is positive, indicating   a long positive tail (many small losses and a few extreme gains), - except for $< ROS >_3$  which has an unexpected negative skewness,  thereby  indicating a long lower range tail (many small gains and several extreme losses).

Next consider whether some "cause" suggest why there are such outliers: the histogram for the (stacked) TTA  variables, in  2006 and  2007, is displayed in Fig.1. The major companies are pointed out. The (up or down) order of investments can be better observed on Fig. 2 that, in 45 cases, there was an increase in TTA , i.e., TTA06 $<$ TTA07, and (of course) 17 cases are such that there was a decrease in TTA : TTA06 $>$ TTA07. It seems relevant to distinguish between these two categories in the discussion of features, - see below. The TTA  magnitudes and the two types of investment classes can be distinguished in Fig. 2. The statistical variations are not large, but not negligible.

 The (rounded) statistical characteristics  for  the  12 statistical indicators distributions, hereby considered as a measure  of the growth and profitability  for the 62 STAR companies;   
  are given in  Table \ref{NewTablestatSKSK}. N.B. data in the center of the table is given in per cents;  the skewness and kurtosis are dimensionless scalars. The standard  confidence intervals limits are readily deduced and reported in Table \ref{TableCI}.

The performance efficiency ratios of the 62 companies are not given, for space savings, but those of the outliers, i.e. when the SME having efficiency values fall outside the relevant   $]$$\mu-2\sigma$, $\mu+2\sigma$$[$   interval are listed in Table \ref{Table4outliers}. There are 3 SMEs which are, rather systematically, positive outliers: (58) Terrienergia, (11) Buongiorno, (13) Cairo Communications, and 1 SME which is systematically "negative outlier": (45) Mondo TV. For completeness,  we also  display, in Table \ref{Table4outliers}, the corresponding values for such companies, even when they are not  true outliers in a statistical sense. It is found that all of these are usually close to the end of the  statistical confidence interval; see Table \ref{TableCI}. This is particularly the case of Mondo TV,  for which  all  efficiency ratios, except for those involving 
 $< DS >_3$, are negative. However,  we repeat: such values  almost  fall within the statistical error bars deduced for the whole 69 firm set.

Interestingly, (11) Buongiorno appears most of the times in the top brackets, but appears at the bottom (the worst) for ratios involving $< DA >_3$. Another interesting finding concerns Buongiorno which appears as "almost a negative outlier" in three efficiency ratios; see Table \ref{TableCI} and  \ref{Table4outliers}. On the other hand, (58) Terrienergia and (13) Cairo Communications have very dissimilar performance efficiency behaviors: the former performing better for "growth", the latter performing better for "profitability". Due to the presence of such outliers, it is of course ridiculous to attempt a regression-like study. The resulting coefficients are all pointing to a valid null hypothesis. Nevertheless, it should occur to the reader that those 4 companies are those with the lowest TTA; see Fig. 2. Moreover, Mondo TV is the only one among the outliers which has a TTA06  lower than its TTA07, - this SME had about a 50\% decrease in investment before the crisis. In contrast, Terrienergia, Buongiorno, and Cairo Communications have relatively the highest increases in TTA .

%Notice that one should not be obsessed by the names of the  firms.  They serve for pointing to statistical values and their consequences. 
%The statistical findings  "speak by themselves". 

Results of correlations can be illustrated through figures, on which the highest TTA  firms are more easily distinguished. However, in view of the above and Table 2, it should occur to the reader that such companies had not a  well performing strategy. Indeed, a few of these "not systematically outlier companies" have a mixture of positive (or negative) small efficiency ratio values. One should observe that

 \begin{figure}%3
  \begin{center}%centering
\includegraphics[scale=.65] {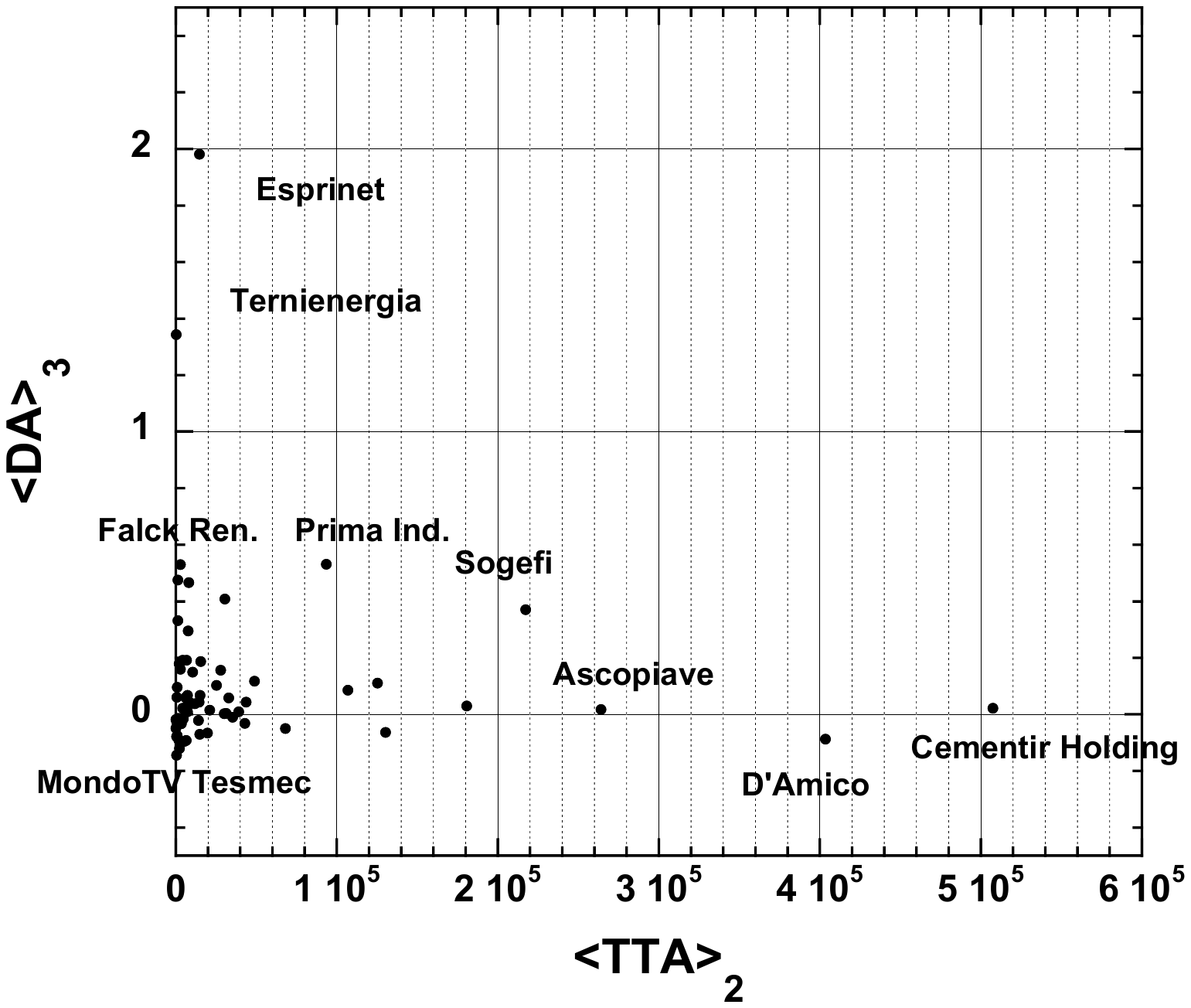}
\caption{  Relation $<DA>_3$   $vs.$ $<TTA>_2$   for the 62 companies listed on the 
STAR market: observe a large $<DA>_3$ effect for Esprinet and Ternienergia without 
much $<TTA>_2$;  a small negative $<DA>_3$ with large $<TTA>_2$ for  D'Amico and slightly positive $<DA>_3$ 
with large $<TTA>_2$ for Cementir Holding and Ascoclave.} 
\label{Plot52DAssets3TTA2lili.pdf}
\end{center}
\end{figure}

 \begin{figure} %4
  \begin{center}%centering
\includegraphics[scale=.65] {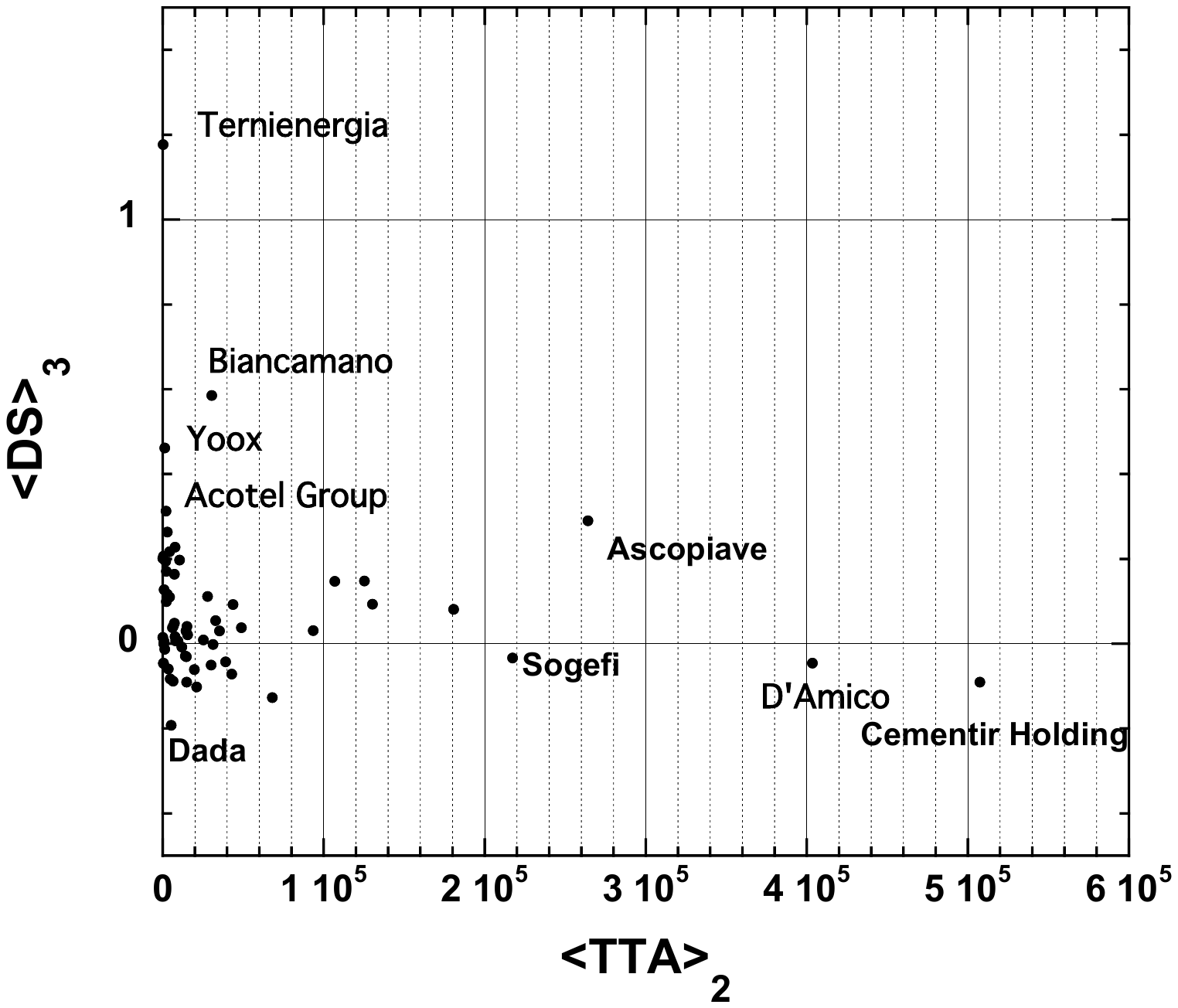}
\caption{  Relation $<DS>_3$   $vs.$ $<TTA>_2$   for the 62 companies listed on the STAR market: observe a large $<DS>_3$ effect for Ternienergia without much $<TTA>_2$; a negative $<DS>_3$ with  large $<TTA>_2$ for D'Amico and Cementir Holding
.} 
\label{Plot51DSales3TTA2lili.pdf}
\end{center}
\end{figure}

\begin{figure} %5
  \begin{center}%centering
\includegraphics[scale=.65] {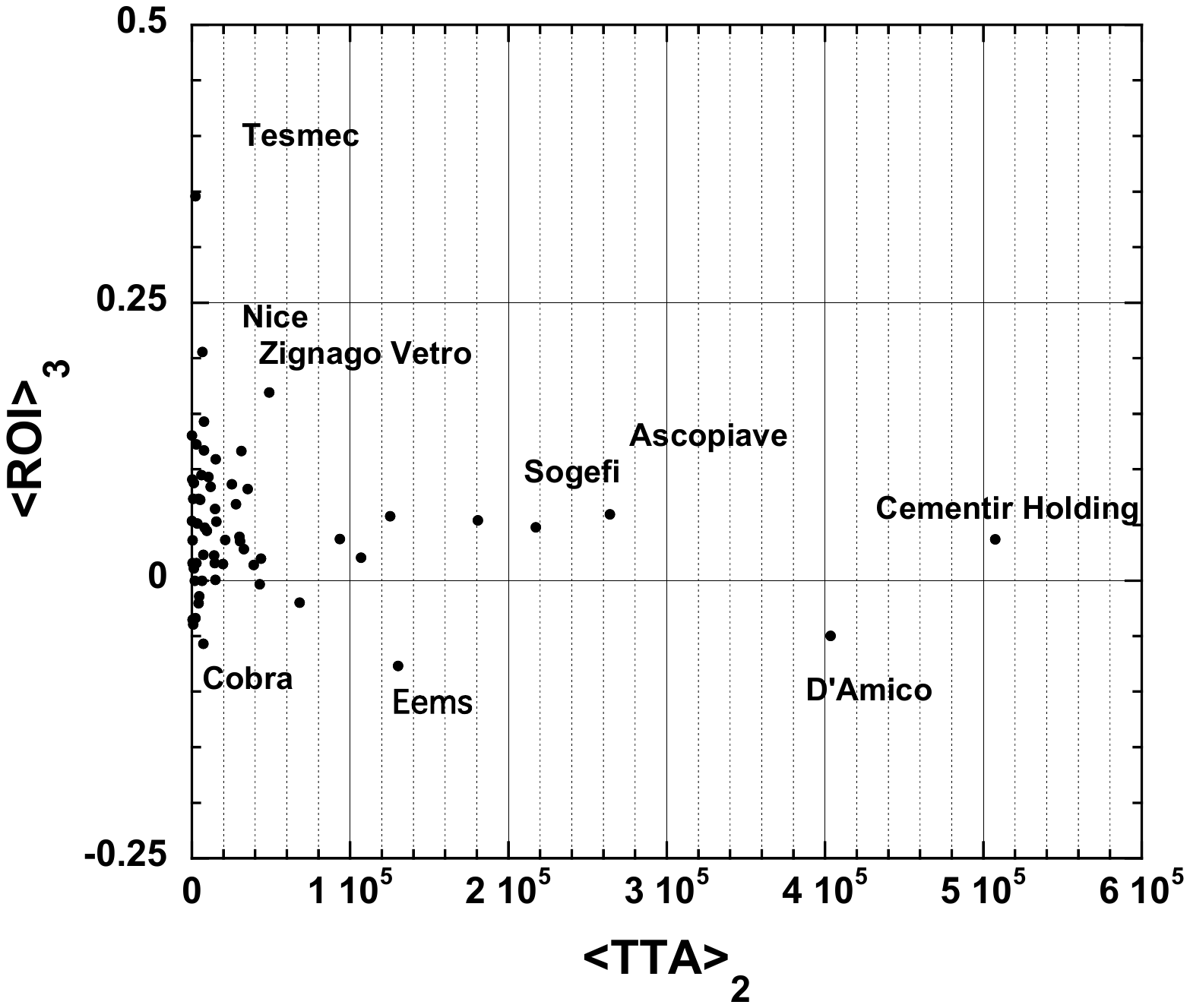}
\caption{  Relation $<ROI>_3$   $vs.$ $<TTA>_2$   for the 62 companies listed on the STAR market:  observe a weak$<ROI>_3$ effect  with large $<TTA>_2$    for Cementir Holding and Ascopiave; a much larger but negative  for  D'Amico;  a large $<ROI>_3$ with low$<TTA>_2$  for  Tesmec; the  negative largest  $<ROI>_3$ for Eems.} 
\label{Plot54ROI3TTA2lili.pdf}
\end{center}
\end{figure}

\begin{figure} %6
  \begin{center}%centering
\includegraphics[scale=.65] {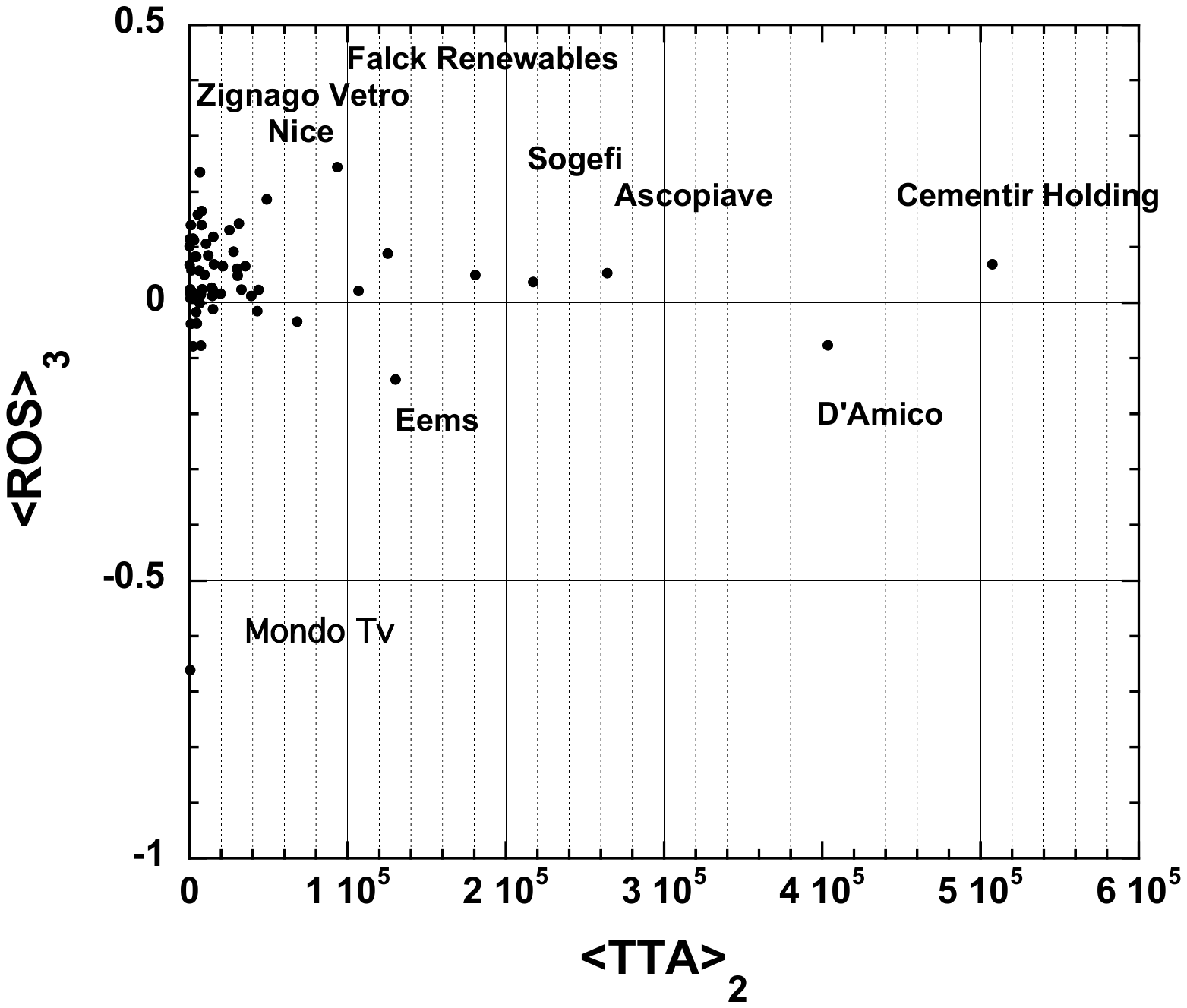}
\caption{  Relation $<ROS>_3$   $vs.$ $<TTA>_2$   for the 62 companies listed on the STAR market:  see a moderate $<ROS>_3$ effect for Sogefi,  Ascopiave, D'Amico and     Cementir Holding; a large negative  $<ROS>_3$ effect for Mondo TV; the best $<ROS>_3$ effect for  Falck Renewables, Zignago Vetro, and     Nice.} 
\label{Plot55ROS3TTA2lili.pdf}
\end{center}
\end{figure}

 \begin{itemize}
 \item  • Fig. 3 displays the relationship between $< DA >_3$  and $<TTA>_2$; the largest $< DA >_3$  effect occurs for Esprinet and Ternienergia, both with a low $<TTA>_2$. A small negative  $< DA >_3$  for D’Amico which has a large $<TTA>_2$ is observed, in contrast to Cementir Holding and Ascopiave which have a large $<TTA>_2$ also, but with a slightly positive $< DA >_3$;

\item • Fig. 4 displays the relationship between $< DS >_3$ and  $<TTA>_2$: a large $< DS >_3$  effect occurs for Ternienergia (recall that it has a low $<TTA>_2$, as already emphasized); a negative $< DS >_3$ effect occurs for D'Amico and Cementir Holding;

\item  • Fig. 5 displays the relationship between $< ROI >_3$ and $<TTA>_2$: a weak $< ROI >_3$  effect is found for Cementir Holding and Ascopiave; a negative but much larger occurs for D'Amico; in contrast, a large $< ROI >_3$ occurs for Tesmec, while the negatively largest  $< ROI >_3$   is for Eems, - both firms with rather low $<TTA>_2$; 

\item • Fig. 6 displays the relationship between $< ROS >_3$ and $<TTA>_2$; a moderate $< ROS >_3$ positive effect occurs for Sogefi, Ascopiave, D'Amico and Cementir Holding, the four largest TTA companies; a large negative $< ROS >_3$ effect occurs for Mondo TV; on the opposite side, the best $< ROS >_3$ positive effect is for Falck Renewables, Zignago Vetro, and Nice.
 \end{itemize}

\section{Conclusion} \label{conclusion}
 In this paper, we have proposed  arguments in favor of extreme values to shine light on performance checking. 
The fact that a comprehensive set of outlier properties can be derived for  measured "anomalous performance ratios"  is a considerable
attraction. Recall that these include two aims presently envisaged. Finding a convenient measure of investment performance, - whatever the investment, and from such a measure observe at the time of crisis what positive or negative effect has an investment "cause". The key timing separating the cause and its effect is the financial crisis. Notice that the study allows three considerations from extreme value  analysis: not only the investment evolution; up or down, low or high, but also through their average, serving as a control kind of test. It should be obvious that the best performance should be better appreciated when (unexpectedly?)  the investment is low. This has been emphasized through Table 2.

For further  introducing  the following discussion, let us  briefly define,  in Table \ref{TableProvATINinhabNc}
  the type of companies mentioned here above. Observe that they cover various sectors. This allows us to consider that the 62 STAR market companies represent an interesting and valuable set of SMEs for our investigation. 
Let it be  observed that the positive outliers belong to different activities: Terrienergia: Utilities; Buongiorno: Technology; Cairo Communications: Media, while the negative outlier Mondo TV is also a Media actor. Therefore, a "SME segment independent universal rule" is found: all those 4 companies have the lowest TTA  of the STAR market; recall Fig. 2.
 
 Nevertheless, there are differences: Terrienergia and Cairo Communications have very dissimilar performance efficiency behaviors, the former performing better for "growth", the latter for "profitability". Since Terrienergia, Buongiorno, and Cairo Communications have a high increase in TTA, one might recommend such a strategy. In fact, Mondo TV did not increase its TTA, pointing to a deficient strategy, - again pointing that the timing of "investment" seems relevant; not the average value.

 Conclusions follow, expecting that SME  board members 
%practitioners   
understand statistical facts about successful and unsuccessful strategies. % Let those be stressed. 
 
 Specifically, one of the targets of the present paper has been to demonstrate (and discuss) the effect of a cause (found to be the assets) on the  statistical mean of performance variables (either growth or profitability). 
 
 %assets (TTA). The variables or financial/economic indicators (i  = 1, … , 4 ) representing business performance are "growth", expressed in terms of (1) sales variations (DS), (2) total assets variations (DA), and "profitability" through (3) returns on investments (ROI) or (4) returns on sales (ROS). 
 
{\it In fine}, observe that this paper has been arguing that it is "truly interesting" to look at the extremes  in distribution tails, indeed (Caroni and Karioti, 2004; Unnikrishnan, 2010; Gumedze and Chatora, 2014). The fact that a comprehensive set of theoretical properties can be derived from extreme values  is a considerable attraction. Even more important, from the practical point of view, is that our results provide  more evidence on   deducing some new principle, - here in order to optimize strategies within a forecasting perspective (Caldeira et al., 2016; Kapetanios et al., 2016), but surely in other cases as well.  % (Hill, 1975)
  The statistical analysis in this paper points  that the best resistance  to crisis  is found to occur for the firms with the initially lowest assets.   An extremely important point resides in the timing of investments. An increase leads to a much better performance than a decrease starting from a higher level. We conclude that applied statistics  studies are relevant in the context of optimizing some performance strategy.
 
% It seems interesting to know that the relatively smallest investment performs the best, as long as there is a strategic TTA  increase.
  \clearpage

 \begin{table}
  \begin{center} \begin{tabular}{|c|c|c|}
\hline
i:	&	Name	&	"Super sector"	\\ \hline						
1	&	Acotel Group	&	Telecommunications\\
5	&	Ascopiave	&	Utilities	\\
7	&	Biancamano & Industrial goods \& Services	\\						
11	&	Buongiorno	$^*$&	Technology  \\						
13	&	Cairo Communication & Media	\\						
15	&	Cementir	Holding&	 Constructions \& Materials	\\						
17	&	Cobra	  &	 Industrial goods \& Services	\\
18	&	Dada	  &	 Industrial goods \& Services	\\
20	&	D'Amico	&	Industrial goods \& Services	\\						
25	&	Eems	$^{**}$&	Technology  	\\						
30	&	Esprinet	&	Technology	\\
31	&	Eurotech	&	 Technology	\\						
33	&	Falck Renewables &	 Utilities	\\						
45	&	Mondo TV & Media	\\						
46	&	Nice	&	Industrial goods \& Materials	\\	
50	&	Prima Industrie&	 Industrial goods \& Materials	\\						
57	&	Sogefi	&	Automobiles \& Parts	\\						
58	&	Ternienergia	&	Utilities	\\						
59	&	Tesmec	&	Industrial goods \& Services	\\						
61	&	Yoox	 $^{***}$&	Retailer  	\\						
62	&	Zignago Vetro	&	Industrial goods \& Services	\\\hline						
\end{tabular}  \end{center}
\caption{  A few STAR company names which are mentioned in the text, or in figures, in alphabetical order (index $i$), and their business type.} N.B.
* Since July  2012, Buongiorno is part of Docomo Digital \\
** Eems was moved away from Technology in the STAR to the MTA Market/Segment\\
*** In March  2015, Yoox merged with Net-a-Porter  \label{TableProvATINinhabNc}
  \end{table}

\end{document}